\documentclass[showpacs,pre,twocolumn,superscriptaddress]{revtex4-1}

\usepackage{amsmath,amssymb}
\usepackage{graphicx}
\usepackage{dcolumn}
\usepackage{bm}
\usepackage{color}
\usepackage[normalem]{ulem}

\begin{document}
\title{The consequences of dependence between the formal area efficiency and the macroscopic electric field  on linearity behavior in Fowler-Nordheim plots}

\date{\today}

\author{Thiago A. de Assis}
\address{Instituto de F\'{\i}sica, Universidade Federal da Bahia,
   Campus Universit\'{a}rio da Federa\c c\~ao,
   Rua Bar\~{a}o de Jeremoabo s/n,
40170-115, Salvador, BA, Brazil}
\email{thiagoaa@ufba.br}

\author{Fernando F. Dall'Agnol}
\address{Universidade Federal de Santa Catarina,
 Campus Blumenau,
   Rua Pomerode 710 Salto do Norte, 89065-300,
Blumenau, SC, Brazil}
\email{fernando.dallagnol@ufsc.br}

\author{Roberto F. S. Andrade}
\address{Instituto de F\'{\i}sica, Universidade Federal da Bahia,
   Campus Universit\'{a}rio da Federa\c c\~ao,
   Rua Bar\~{a}o de Jeremoabo s/n,
40170-115, Salvador, BA, Brazil}
\email{randrade@ufba.br}

\begin{abstract}

%small changes
This work presents a theoretical explanation for a crossover in the linear
behavior in Fowler-Nordheim (FN) plots based on cold field electron emission
(CFE) experimental data. It is characterized by a clear change in the decay
rate of usually single-slope FN plots, and has been reported when non-uniform
nano-emitters are subject  to high macroscopic electric field $F_M$. We assume
that the number of emitting spots, which defines an apparent formal area
efficiency of CFE surfaces, depends on the macroscopic electric field.
Non-uniformity is described by local enhancement factors
$\left\{\gamma_j\right\}$, which are randomly assigned to each distinct emitter
of a conducting CFE surface, from a discrete probability distribution
$\rho{(\gamma_{j})}$, with $j=1,2$. It is assumed that $\rho{(\gamma_{1})} <
\rho{(\gamma_{2})}$, and that  $\gamma_{1} > \gamma_{2}$. The local current
density is evaluated by considering a usual Schottky-Nordheim barrier. The
results reproduce the two distinct slope regimes in FN plots when $F_M \in $
$[2,20]$ V/$\mu$m and are analyzed by taking into account the apparent formal
area efficiency, the distribution $\rho$, and the slopes in the corresponding
FN plot. Finally, we remark that our results from numerical solution of
Laplace's equation, for an array of conducting nano-emitters with uniform apex
radii $50$ nm but different local height, supports our theoretical assumptions
and could used in orthodox CFE experiments to test our predictions.

\end{abstract}

\maketitle

\section{Introduction}

%small changes
Understanding the role of the morphology of large area field electron emitters
(LAFEs) is of utmost importance  to better explore their potential
applications. Typical field emitter arrays consist of regular two-dimensional
patterns of individual, similar, and small size field electron emitters, which
may be prepared by lithographic techniques \cite{MC}. The best known LAFE
devices are the Spindt arrays, in which each individual field emitter is a
small sharp molybdenum cone \cite{Spindt}. Unfortunately, there are some
inconveniences of using Spindt-type metal arrays for vacuum microelectronic
devices due the expensive production, the critical lifetime in technical vacuum
and the high operating voltages \cite{Oliver}. Moreover, poor tip-to-tip
reproducibility caused by the presence of nano-protrusions, which are also
present in other nonmetallic arrays, makes it difficult to accurately predict
their emission characteristics. To sidestep some of these difficulties, the
cold field emission (CFE) community redirected efforts to study and produce
different purpose LAFEs as nano-electronic devices, including carbon
nano-structures which have near-ideal whisker-like shapes with hemispherical
tips \cite{Cole}. This choice is justified by a set of favorable properties
like nanometer size tip, high chemical inertness, high electrical and thermal
conductivity, and low manufacturing costs  \cite{Oliver}.

A relevant issue relating experimental and theoretical aspects of CFE studies
is how to assess, with sufficient technologic reliability, several quantities
related to the LAFE efficiency  from measurable current-voltage
characteristics. This is usually done using Fowler-Nordheim (FN) plots, which
relates the macroscopic current density $J_{M}$ to the applied (or macroscopic)
electric field $F_{M}$.  The theory leading to Fowler-Nordheim-type (FN-type)
equations suggests to draw FN-plots consisting of curves for
$\ln\{J_{M}/F_{M}^{2}\}$ vs $1/F_{M}$, but other variable combinations can be
used as well (see for instance Ref. \cite{MC}). Actually, FN-plots may present
a non-linear behavior and is necessary to set up a convenient theory that takes
into account more realistic conditions under which a specific CFE experiment is
performed in order to obtain a correct interpretation of the field enhancement
factor (FEF) and other experimental outputs \cite{Cahay}. In this context, it's
important to discuss some general definitions as follow: the slope
characterization parameter (alternatively called apparent FEF) is defined by

\begin{equation}
\beta^{app} = -\frac{b\phi^{3/2}}{S^{fit}},
\label{Eq1rev}
\end{equation}
where $S^{fit}$ is the slope of a sufficient linear FN-plot, for a given range
of $F_{M}$, $\phi$ is the local work-function of the emitter, and $b$ is the
second Fowler-Nordheim (FN) constant $(\approx 6.830 890$ eV$^{-3/2}$ V
nm$^{-1}$); the actual characteristic FEF, $\gamma_{C}$, is defined as

\begin{equation}
\gamma_{C} = \frac{F_{C}}{F_{M}},
\label{Eq2rev}
\end{equation}
where $F_{C}$ is the characteristic local barrier field. Then, the general
relationship between $\gamma_{C}$ and $\beta^{app}$ has the form

\begin{equation}
\gamma_{C} = \sigma_{t} \beta^{app},
\label{Eq2rev}
\end{equation}
where $\sigma_{t}$ is the relevant generalized slope correction factor.

Some situations can display nonlinear behavior in the corresponding FN-plots.
This can be observed already in the pioneer work by Lauritsen who, in this Ph.
D. thesis obtained plots of the form
$\log{(i_e)}$ vs $1/$Voltage, where $i_e$ is the macroscopic current emitted.
He found experimentally that plots of the form $\log{(i_e)}$ vs $1/$Voltage may
be consisted of two straight lines, with a slight kink in the middle, using a
cylindrical wire geometry \cite{Lauritsen} (see, for instance, Figs. 6 and 12
of that work). Another example is related to the particular condition in which
a large series resistance is found in the circuit between the high-voltage
generator and the emitter's regions. The interpretation of corresponding
FN-plots was provided by Forbes and collaborators \cite{FNReport}. For both
LAFE and single tip field emitters (STFEs), they showed that if the so-called
CFE orthodox emission hypotheses \cite{Forbes1} are not satisfied, the analysis
of the results based on the elementary FN equation, as usually performed by
experimentalists, can generate a spurious estimates for the true electrostatic
FEF \cite{AuPost,Forbes1}. On the other hand, recent theoretical works by one
of authors \cite{deAssis12015,deAssis22015} explained how a slight positive
curvature on FN-plots arises when a dependency between the apparent formal area
efficiency ($\alpha_{f}$) and $F_M$ is taken into account. For some assumptions
of non-uniform conditions in the LAFES morphology, which amounts to consider a
local FEF ($\gamma$) probability distribution $\rho(\gamma)$ with exponential
or Gaussian behavior, the orthodoxy test showed does not fail for practical
circumstances. Despite this, it was possible to suggest experimental tests that
can verify the proposed correction to the $\beta^{app}$ values with statistical
significance.

%small changes
In this work, the authors investigate the conditions under which a clear
crossover on the FN plots of CFE may appear, by assuming that it is only a
consequence of the dependency between $\alpha_{f}$ and $F_{M}$. The electron
emission from a conduction band on a particular LAFE location is described by
FN-type equations with a Schottky--Nordheim (SN) barrier. Different from Refs.
\cite{deAssis12015,deAssis22015}, which considered continuous $\gamma$
distributions, the present model assumes CFE through a non-uniform distribution
of the local FEF $\gamma_{j}$ on LAFE surface, which is described by a discrete
asymmetric bimodal distribution for two distinct values $\gamma_{1} $ and $
\gamma_{2}$, with  $\gamma_{1}>\gamma_{2} $ and $\rho{(\gamma_{1})}
< \rho{(\gamma_{2})}$. So, let us define

\begin{eqnarray}
\label{Eq9}
q &=& \frac{\gamma_{2}}{\gamma_{1}},
\end{eqnarray}
and
\begin{eqnarray}
\label{Eq9rev}
r &=& \frac{\rho(\gamma_2)}{\rho(\gamma_1)}.
\end{eqnarray}
The characteristic FEF of the LAFE is $\gamma_{1}$. From
now on, whenever we mention this specific model we will indicate the
characteristic FEF as $\gamma_{1}$, while $\gamma_{C}$ will be used to refer to
FEF in general conditions. Depending on the bimodal asymmetry parameter
$r\equiv \rho(\gamma_{2})/\rho(\gamma_{1})$, this contribution may lead  to a
clear crossover effect in the corresponding FN plots. Our results suggest that
this simple mechanism, mimicking fluctuations of the individual emitter
morphology on a LAFE surface, can justify a pronounced change in FN plots only
as the emission is orthodox.

This paper is organized as follows. In Sec. \ref{sec.II}, the model and the
equations for computing the local current density $J_L$ are presented. We put
this in perspective of previous studies discussing nonlinear behavior in the
corresponding current-voltage measurements. Results are  presented in Sec.
\ref{modelres}, focusing on the conditions where nonlinear FN plots can be
found. We also discuss the results from numerical solution of Laplace's
equation, using an array of conducting nano-emitters with large apex radii
($50$ nm) but different heights. In Sec. \ref{conc}, the main conclusions are
presented.

\section{Current density calculations, model and previous works}
\label{sec.II}

The interpretation of experimental CFE outputs have often been done using the
elementary FN-type equation, hereafter referred to as ``elementary" equations
and theory, which considers the quantum-mechanical electron tunneling across an
triangular barrier. However, it known since the 1950's that this equation
under-predicts current density by a factor of $10^{2}$ to $10^{3}$
\cite{Forbesreformulation}, specially in the case of bulk metals. A physically
complete FN-type equation \cite{Forbes2008} for the local current density $J_L$
can be written as

%\begin{equation}
%J_L = \lambda^{GB}_{L} a \phi^{-1} F_{L}^{2}  \exp{\left(-\nu^{GB}_{F} b %\phi^{3/2}/F_{L}\right)},
%\label{Eq1}
%\end{equation}

\begin{equation}
J_L = \lambda_{L} a \phi^{-1} F_{L}^{2}  \exp{\left(-\nu b \phi^{3/2}/F_{L}\right)}.
\label{Eq1}
\end{equation}
Here, $\nu$ is the  barrier form correction factor associated with barrier
shape, and $\lambda_{L}$ takes into account all other effects, including
electronic structure, temperature, and corrections associated with integration
over electronic states. In this work, we are restricted to the tunneling of
electrons close to the Fermi level, so that we implicitly assume that $\nu$
takes into account this fact, and we refrain from explicitly adding a subscript
``$F$" to $\nu$. $a (\approx 1.541434 \times 10^{-6}$ A eV V$^{-2}$) and $b$
(the latter defined in Introduction) are the first and second Fowler-Nordheim
(FN) constants, respectively, while $\phi$ is the local work function and
$F_{L}$ is the local electric field.

The correction associated with \textcolor[rgb]{0.00,0.00,1.00}{a} SN barrier
(used in Murphy-Good theory \cite{Murphy}), which accounts for the potential
energy contribution resulting from the interaction of the electron with its
image charge, is written as \cite{Forbesreformulation,Forbes2}

\begin{equation}
\nu^{SN} \approx 1-\textit{f}+(1/6)\textit{f}\ln(\textit{f}),
\label{Eq2}
\end{equation}
where $\textit{f} \equiv F_{L}/F_{R}$. Since $F_{R}\equiv e^{3}/\left(4 \pi
\epsilon_{0} \phi^{2}\right)$, where ``$e$" is the positive elementary charge
and $\epsilon_{0}$ is the electric constant, is the value of the external field
for which height of the tunneling barrier vanishes, $f$ represents the scaled
value of $F_L$. It plays a relevant role in CFE theory as a reliable criterion
to test if the emission is orthodox or not \cite{Forbes3}. Indeed, from a FN
plot based on data points, it's possible to derive values for $f^{extr}$
\cite{Forbes3,Forbes1} from the equation
 %``\textit{f}-values"

\begin{equation}
f^{extr} = -\frac{s_{t} \eta(\phi)}{S^{fit} \left(1/F_{M}^{exp}\right)}.
\label{Eq3}
\end{equation}
If orthodox emission hypothesis is respected, all independent variables are linearly related to each other, and ``$f$" can be used as a scaled value of the variable ``$F_{L}$" \cite{Forbes1}. Then, in data analysis based on the orthodox emission hypothesis, Eq.(\ref{Eq3}) applies for all appropriate choices of independent and dependent variables and guarantees that the test for lack of orthodoxy works for any physically relevant form of FN plot.
Let us remark that all quantities in Eq. (\ref{Eq3}) are directly accessible
from CFE experiments or have been previously obtained for typical conditions in
conductor materials \cite{Trolan}. The parameter $\eta(\phi) \equiv b
\phi^{3/2}/F_{R}$ depends only on the work-function $\phi$, while $S^{fit}$ is
the slope of a sufficient linear FN-plot for a given range of the macroscopic
electric field. The symbol $s_{t}$ represents
the ``fitting value" of the slope correction function for the SN barrier, and
can be approximated by $\approx 0.95$. It plays a similar role to the symbol
$\sigma_{t}$ in Eq. (\ref{Eq2rev}) and, since we restrict our work to SN
barriers, it will replace $\sigma_{t}$ from now on.  Equation (\ref{Eq3})
provides estimates of the values of $f^{extr}$ that correspond to
macroscopic-field values apparently inferred from experiment.

In this work, we constructed FN plots of the form $\ln\{J_{M}/F_{M}^{2}\}$ vs
$1/F_{M}$. If the emission is orthodox, it's possible to measure directly the
values of $\gamma_{C}$, once the characteristic point ``C" over a LAFE device
is defined as apex of the structure, representing the tip with the highest apex
field.

Over an experimental LAFE surface, it is possible to find an almost continuous
distribution of local $\gamma$ values. However, considering two most prominent
emitting locations on LAFE, it is convenient to approximate such a distribution
by a discrete one, with at most two distinct values of $\gamma_j$ (j=1,2),
namely $\{\gamma_{1}=\gamma_{C},\gamma_{2}\}$, so that  $\rho{(\gamma_1)} +
\rho{(\gamma_2)}  = 1$  with $\gamma_{1} > \gamma_{2}$. Therefore, as already
mentioned, our analysis is restricted to a bimodal distribution for the local
FEFs of LAFE emitters. Indeed, any other
location in the LAFE will be considered as having a FEF $\gamma_3\ll\gamma_2$.
Under this assumption, the corresponding local current density
$J_{L}^{3}\approx0$ so that we can restrict all following expressions to the
values $j=1$ and $2$.

Using Eqs.(\ref{Eq1}) and (\ref{Eq2}), it is possible to write an expression
for the site $j$ dependent local current density $J_{L}^{j}$ in a LAFE surface
(see Refs. \cite{deAssis22015} and \cite{ForbesAPL}) under the assumption of a
SN barrier as

\begin{multline}
J_{L}^{j}(\phi ,F_{M},\gamma_{j}) = \lambda_{L} a \phi^{-1} \exp{\left[\eta(\phi)\right]} F_{R}^{\eta(\phi)/6} (\gamma_{j} F_{M})^{\kappa} \times \\
\times \exp{\left[-b \phi^{3/2} / (\gamma_{j} F_{M})\right]},
\label{Eq4}
\end{multline}
where $\kappa \equiv 2 - \eta(\phi)/6$, the local field $F_L$ is replaced by
$\gamma_j F_M$, and $F_M$ lies in the range 2 V/ $\mu$m $\leq F_M \leq$ 20
V/$\mu$m, which are the typical conditions for CFE technologies that use
nano-sized diameters. We remark that, depending on the barrier shape,
$\lambda_{L}$ can assume values over a wide interval $0.005< \lambda_{L} <11$
\cite{FNReport}. In this work, we always consider $\lambda_{L}=1$.

\begin{figure}
\includegraphics [width=10.0cm,height=8cm] {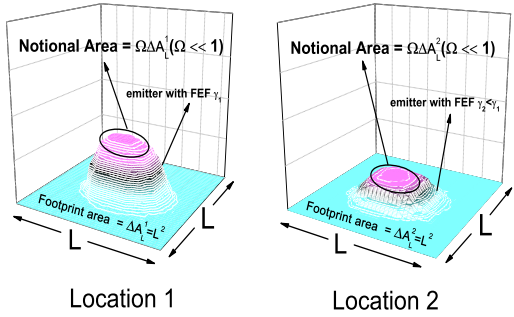}
\caption{(Color online) Illustration of the single tips used in a LAFE with $j=1$ (left) and $j=2$ (right) (locations of an array of nanostructures) and the corresponding footprint of areas ($\Delta A^{j}_{L}$). The related notional emission area ($\Omega \Delta A^{j}_{L}$) is also indicated.} \label{ress1}
\end{figure}

Summing up over the possible values of $\gamma_j$, the total $J_M$ current
density is written as

%changes
\begin{equation}
J_{M} = i_{e}/A_{M} = n_{L}  \frac{\sum\limits_{j=1,2} \rho_{j}{(\gamma_j)}
J^{j}_{L}(\phi ,F_{M}, \gamma_{j}) \Omega \Delta A^{j}_{L}}{A_{M}
\sum\limits_{j=1,2} \rho_{j}{(\gamma_j)}},
\label{Eq5}
\end{equation}
where $i_{e}$ is the total emission current, and $\Omega \Delta A^{j}_{L}$
($\Omega$ represents a typical notional area efficiency of a field emitter) is
the notional emission area associated with the $j-$th FEF-value which, in a
first approximation, is considered to be independent of $F_{M}$. This
approximation is very good since, for usual values of $F_{M}$ of the order of
few $V/\mu$m, $\Omega \Delta A^{j}_{L}$ is only weakly dependent of $F_{M}$
(see Sec.\ref{Posts}). Fig.\ref{ress1} shows a representation of the emitters
used in LAFE and the corresponding ``footprint" of areas $L^2$.

We remember that Eq. (\ref{Eq5}) considers negligible the total emission contribution where the FEF is effectively unity, i.e., at planar regions of footprint.
For a plausible estimation of $\Omega$, which is expected to be much less than unit, we consider the following arguments: experimental values of macroscopic current density are often around $10$mA/cm$^2$. However, according to Dyke and Dolan \cite{DykeTrolan}, a mid-range local current density might be around $10^{4}$ A/cm$^2$. This suggest that typical experimental notional area efficiencies might be around $10^{-8}$ (this is confirmed in Sec.\ref{Posts} for our electrostatic simuations with hemispherical tips). Then, in this work, we investigate a device with an array of isolated nanostructures, where $\Omega \approx 10^{-8}$. Finally, the sum in Eq. (\ref{Eq5}) is taken over the macroscopic substrate footprint area of the emitter, $A_M$, which contains a number of locations, $n_{L}$, each one with footprint of area $L^{2}$ as shown in Fig. \ref{ress1}.
The macroscopic current density $J_{M}$ can also be written as:

\begin{equation}
J_M = \alpha_{n} J_{C} = \alpha_{n} \lambda_{C}  J_{kC} = \alpha_{f} J_{kC},
\label{Eq6}
\end{equation}
where $\alpha_{n}$ is the notional area efficiency, $\alpha_{f}$ has already been defined in Section 1, $\lambda_C$ has a similar meaning as that of $\lambda_L$ in Eq. (\ref{Eq1}). In this work, it is assumed that $\lambda_{C}=\lambda_{L}=1$, so that $\alpha_{n} = \alpha_{f}$. Finally, the kernel current density for the (image-force-related) SN barrier is given by

\begin{multline}
J_{kC}(\phi ,F_{M}) = a \phi^{-1} \exp{\left[\eta(\phi)\right]} F_{R}^{\eta(\phi)/6} (\gamma_{C} F_{M})^{\kappa} \times \\
\times \exp{\left[-b \phi^{3/2} / \gamma_{C} F_{M}\right]}.
\label{Eq7}
\end{multline}

In Ref.\cite{deAssis22015}, the dependency between $\alpha_{f}$ and $F_{M}$ was evaluated for for the case in which $\rho(\gamma_{j})$ corresponds to a family of Gaussian distributions, with different values of the variance $\Delta \gamma$.  The results indicated a slight decreasing change in the slope of the FN plot, for large values of $F_M$ and $\Delta \gamma$. These non-linear behavior was not large enough to cause a failure of the orthodoxy test, nor was able to give rise to two $F_M$ intervals with well defined and different slopes. As it will be shown in the next section, the latter may appear in the present model under specific conditions of the bimodal distribution function, which includes the vales of $q$ and $r$.

Nonlinear behavior in FN plots have been reported in several recent CFE experiments \cite{Li,Patra,Ahmed,Zhang,Li2,Lu,JVSTB2003}, where the discussion of their results were based on the elementary FN equation. Moreover, we pondered that some of the results have showed do not pass the orthodoxy test, and cannot to be interpreted only on the light of the results of the present work (which consider only orthodox field emission), despite similar forms of FN plots have been obtained. For instance, in Ref.\cite{Li2} the field emission properties of ``flexible SnO$_{2}$ nanoshuttle" led to FN plots with a clear crossover presenting two quasi-linear sections. As pointed by Forbes \cite{Forbes1}, for both sections, as a consequence of the unorthodoxy emission (possible explanations include field-dependent changes in emitter geometry and/or changes in collective electrostatic screening effects), spurious FEF values have been found.

Ref.\cite{Lu} analyzed the field electron emission properties of well-aligned graphitic nano-cones synthesized on polished silicon wafers. The authors have investigated how the difference between the values of $\gamma_j$ corresponding to two types of emission sites on the LAFE surface affects the effective emission area for a given range of $F_{M}$ values. Unfortunately, some of their experimental outputs have shown also inconsistencies with the orthodox assumptions \cite{Forbes3,Forbes1}. For instance, consider the data shown in Fig. 2 of Ref.\cite{Lu} together with the work function $\phi=5$ eV of graphitic nano-cones. For anodes with diameter $1.5$ mm, $2.0$ mm, $2.5$ mm and $3.0$ mm and low $F_M$ regime (where a sufficient linear FN plot is obtained), we find, respectively, the following corresponding values for the scaled barrier field [see Eq.(\ref{Eq3})]  $f^{extr} \approx$ $0.46, 0.62, 0.79$ and $1.54$. The first value has been found for $1/F_{M}^{exp}$ = $0.06$ $\mu$mV$^{-1}$, while the three further values were found for $1/F_{M}^{exp}$ = $0.0325$ $\mu$mV$^{-1}$. This suggests that, for all cases where non-linear behavior is observed in the corresponding FN plots, a closer investigation is required to provide a  reliable interpretation of the results. In this specific study, this corresponds to the two smaller anodes. Moreover, for the larger anodes with nonuniform substrates, the orthodoxy test clearly fails, despite the linear behavior of the FN plots. Therefore, the corresponding FEFs indicated in these two cases and the corresponding emission areas extracted are questionable. Finally, is important to emphasize that, very recently, Forbes provided a simple confirmation that the SN barrier is a better model for actual conducting emitters than the usual triangular barrier \cite{ForbesIVNC2015} to extract the emission areas. This can be noticed for a tungsten emitter (X89) data from Dyke and Trolan \cite{DykeTrolan} and independent assessment of emitter area made by electron microscopy.

\section{Results and Discussions}
\label{modelres}

\subsection{Formal area efficiency: role of $\rho(\gamma_{1})$ and $q$}
\label{Arho}

Remembering that the formal area efficiency $\alpha_f$ is an experimentally accessible measure of the fraction of the LAFE surface that is actually emitting electrons, let us explicitly indicate its dependency on $F_{M}$ in Eq.(\ref{Eq6}) by writing

\begin{equation}
J_{M} = \alpha_{f}(F_M) J_{kC}.
\label{Eq8}
\end{equation}
After some manipulations using Eqs.(\ref{Eq4}-\ref{Eq6}) and Eqs.(\ref{Eq9}-10), the following expression can be written (see Appendix - A):

\begin{equation}
\alpha_{f}(F_M) = \Omega \rho(\gamma_1) \left\{ 1 + \Gamma{(q,r,\phi, F_{M})}\right\},
\label{Eq11}
\end{equation}
where
%\begin{multline}
%\Gamma{(q,r,\phi, F_{M})} \equiv q^{\kappa} r \times \\
%\times \exp{\left[-b \left(q^{-1} - 1\right) \phi^{3/2} / (\gamma_{1} F_{M})\right]}.
%\label{Eq12}
%\end{multline}
%

\begin{equation}
\Gamma{(q,r,\phi, F_{M})} \equiv q^{\kappa} r \exp{\left[-b \left(q^{-1} - 1\right) \phi^{3/2} / (\gamma_{1} F_{M})\right]}.
\label{Eq12}
\end{equation}
%
% Changes (Abandonei a mencao a gamma_3)
Based on the actual experimental FEF values \cite{Nilson2}, we fix  $\gamma_{1}= 690$, while $\gamma_{2}$ is free to take different values. This is in accordance with the previous assumptions that the active LAFE emission sites fall into two classes, one of which is ``more pointy" than the other, and hence has a higher FEF. Changes in $\gamma_{2}$, with the corresponding changes in $q$, are restricted to the condition that the electric field over the LAFE device does not exceed a few V/nm, while other complicated effects (as destruction of the LAFE device due to thermal effects) have been neglected.

%Figure 1
\begin{figure}
\includegraphics [width=10.0cm,height=7cm] {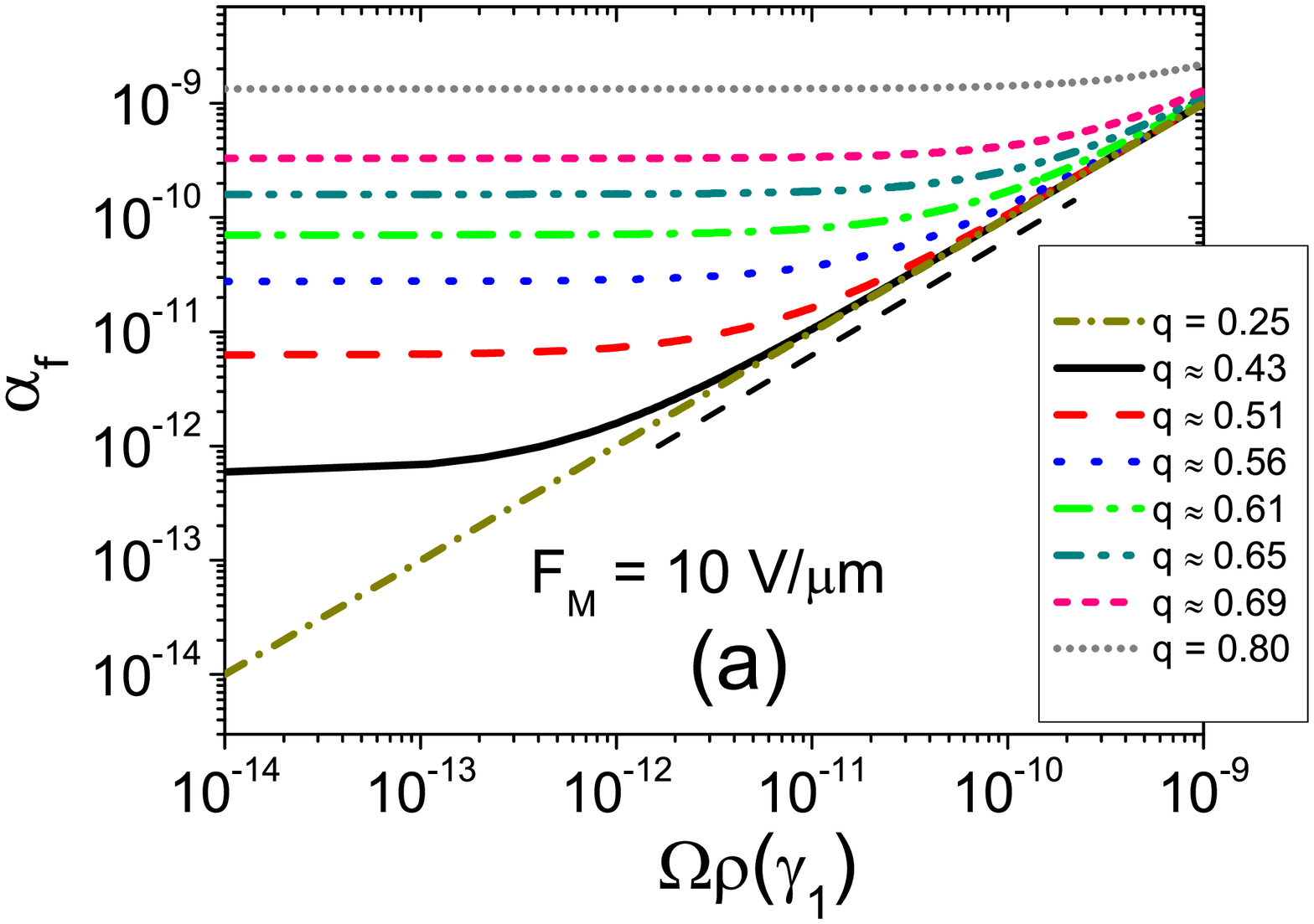}
\includegraphics [width=10.0cm,height=7cm] {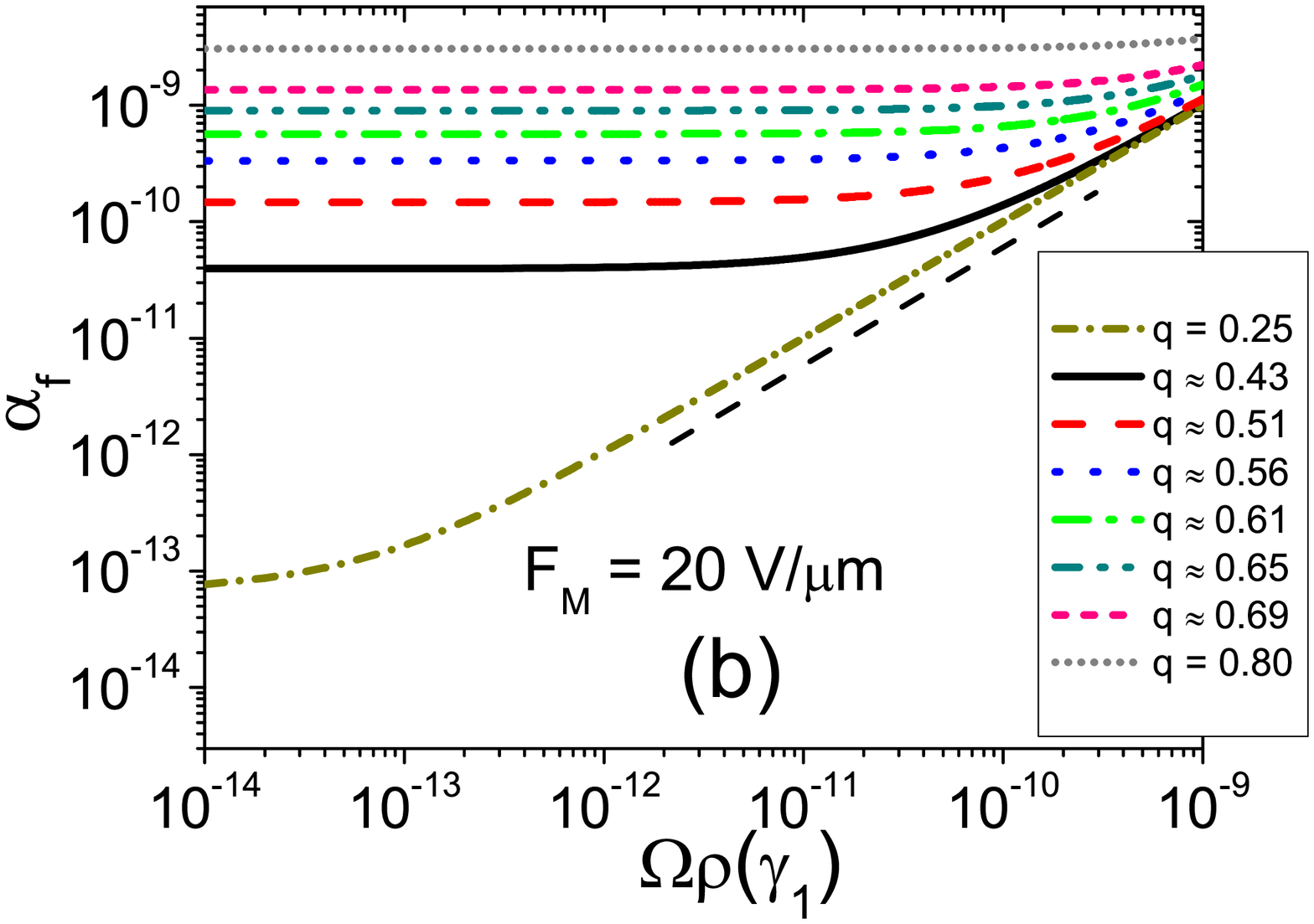}
\caption{(Color online) Behavior of  $\alpha_{f}$ for $10^{-6} \leq \rho(\gamma_1) \leq 10^{-1}$, considering several values of $q$ [see Eq.(\ref{Eq9})] for (a) $F_{M}=10V/\mu$m and (b) $F_{M}=20V/\mu$m. The results are presented for $\Omega = 10^{-8}$ (see text for more details). The dashed (black) lines have slope 1.} \label{Fig1}
\end{figure}
%

%Figure 2
\begin{figure}
\includegraphics [width=10.0cm,height=7cm] {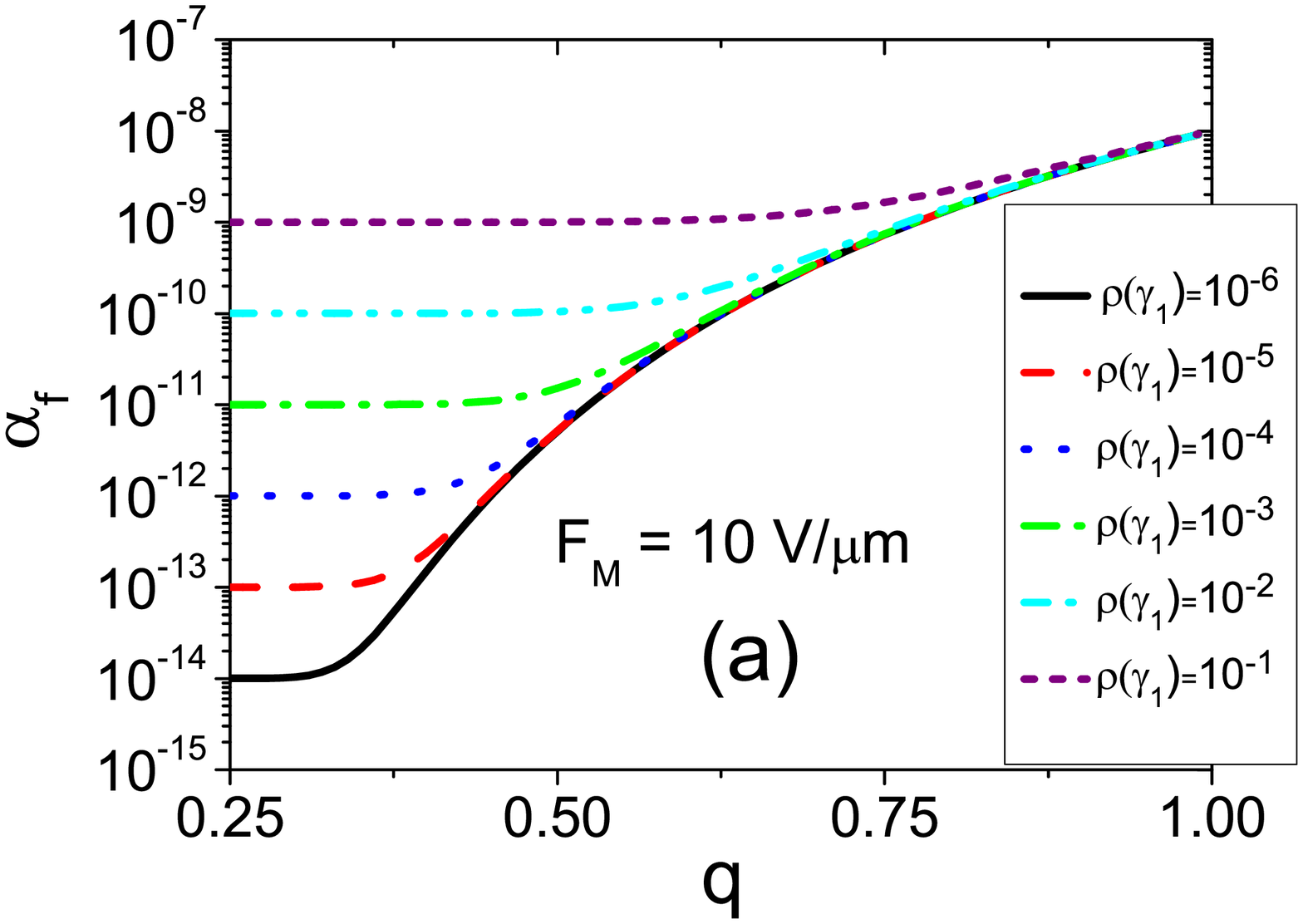}
\includegraphics [width=10.0cm,height=7cm] {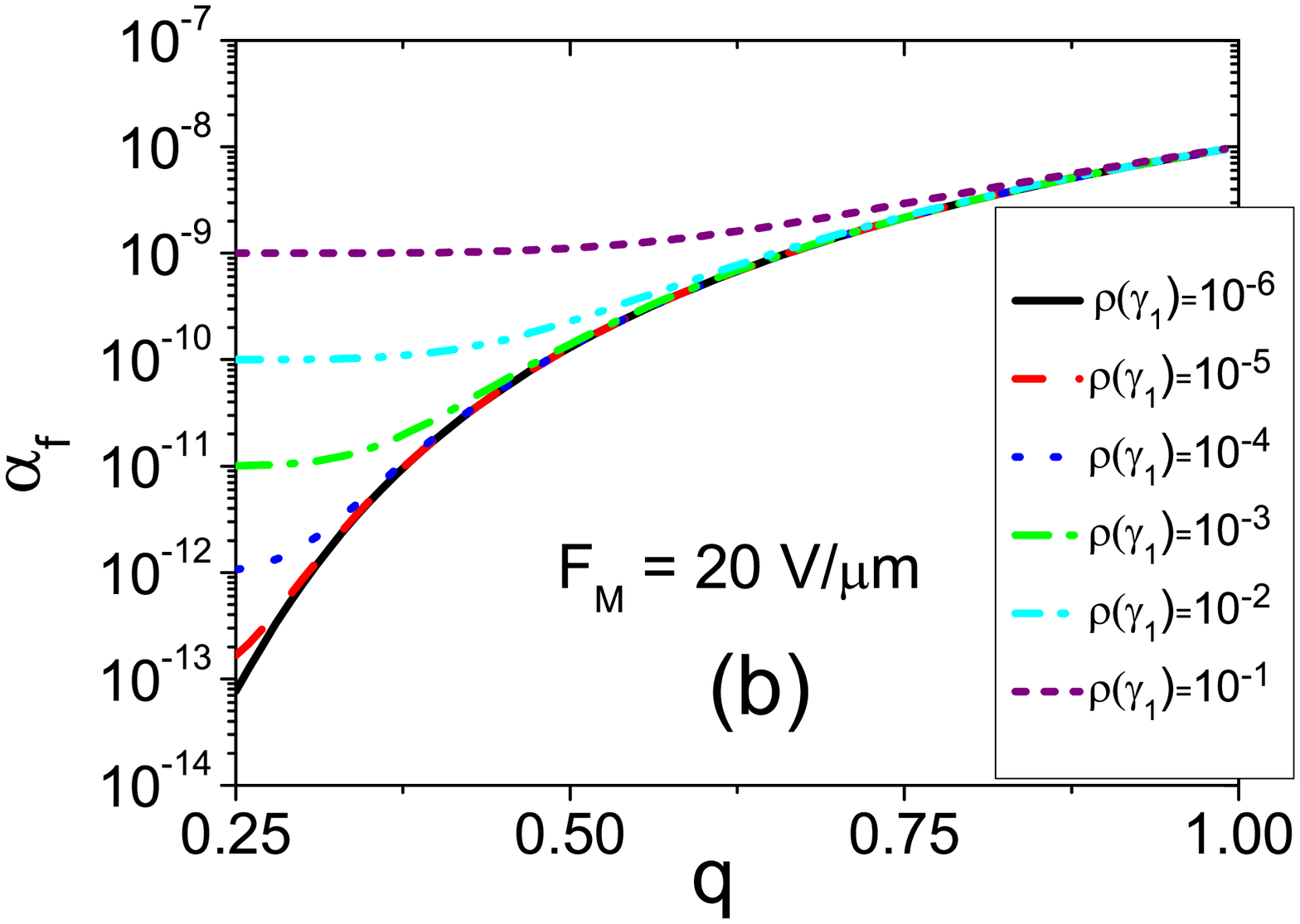}
\caption{(Color online) Behavior of $\alpha_{f}$ for $0.25 \leq q \leq 1$, considering several values of $\rho(\gamma_{1})$ for (a) $F_{M}=10V/\mu$m and (b) $F_{M}=20V/\mu$m. The results are presented for $\Omega = 10^{-8}$ (see text for more details).} \label{Fig2}
\end{figure}

Eq.(\ref{Eq11}) makes it clear that $\alpha_{f}$ depends on $\rho(\gamma_1)$. This is illustrated in Fig. \ref{Fig1}(a) that shows, for several values of $q$ and for a typical value $F_{M}=10V/\mu$m, the behavior of $\alpha_{f}$ as $\rho(\gamma_1)$ changes from $10^{-6}$ to $10^{-1}$. The values of $\alpha_{f}$ were computed by using Eqs.(\ref{Eq11}) and (\ref{Eq12}). For small values of $q$ (e.g., $q \lesssim 0.25$), Fig. \ref{Fig1}(a) shows that $\alpha_{f}$ assumes, approximately, the same values of $\Omega\rho(\gamma_{1})$. In this limit, $\Gamma{(q,r,\phi, F_{M})} \ll 1$ for $F_{M}=10V/\mu$m, and the only emitting spots on the LAFE surface are those with $\gamma_{j}=\gamma_{1}$ for all $10^{-6} \leq \rho(\gamma_1) \leq 10^{-1}$. This behavior is not observed for other values of $q \gtrsim 0.25$ and smaller values of $\rho(\gamma_1)$, when the contribution of the $\gamma_{j}=\gamma_{2}$ regions for the electron emission become relevant as compared with $\gamma_{j}=\gamma_{1}$ regions. However, for larger values of $\rho(\gamma_1)$, again the main emitting spots that contribute to $\alpha_{f}$ are those with $\gamma_{j}=\gamma_{1}$. In this case, the curve bends upwards and $\alpha_{f} \approx \Omega \rho(\gamma_1)$, which is observed as long as $q$ is not so close to 1. Finally, when the limit $q\rightarrow 1$ is approached, the regions with $\gamma_{j}=\gamma_{2}$ contribute to $\alpha_{f}$ for almost all range of values of $\rho(\gamma_1)$. It is important to stress that, as $q$ increases, a more uniform LAFE surface is built, with the presence of second-scale structures presenting close values of $\gamma$. The results shown in Fig. \ref{Fig1}(b) indicate the behavior of $\alpha_f$ at a larger value $F_{M}=20V/\mu$m. In this case, the results suggest that, for values of $q$ close to unity, the regions of the LAFE surface $\gamma_{j}=\gamma_{2}$ also contribute to $\alpha_{f}$ for low values of $\rho(\gamma_{1})$. As will be discussed in the next subsection, when $\alpha_{f} \neq \Omega \rho(\gamma_1)$ and $q$ is not so close to 1, $\alpha_{f}$ depends on $F_{M}$ leading to nonlinear behavior in the corresponding FN plots. Before discussing the behavior of the FN plots, we investigate how $\alpha_{f}$ is related with $q$ when both $\rho(\gamma_{1})$ and $F_{M}$ are kept fixed.

Fig. \ref{Fig2}(a) shows the behavior of $\alpha_f$ as a function of $q$ for several values of $\rho(\gamma_{1})$ and $F_{M}=10V/\mu$m. It's possible to observe that, for higher values of $\rho(\gamma_{1})$, the wider is the interval where $\alpha_{f}$ has a weak dependency on $q$. In this regime, $\alpha_{f} \approx \Omega\rho(\gamma_1)$ and, again, the regions which contributes to $\alpha_{f}$ are only those with $\gamma_{j}=\gamma_{1}$. After the plateau, which increases as $\rho(\gamma_{1})$ increases, $\alpha_{f}$ is expected to depends more strongly on $q$. Fig. \ref{Fig2}(b) illustrate the behavior for $F_{M}=20V/\mu$m. Now the plateau disappears for small values of $\rho(\gamma_{1})$ and, in this regime, $\alpha_{f}$ depends on $q$ in the entire displayed range. For larger values of $\rho(\gamma_{1})$, e.g. $\rho(\gamma_{1})\gtrsim 10^{-2}$, the plateau region is restored. However, even in this range of $\rho(\gamma_{1})$, it's possible to observe the weak dependency between $\alpha_{f}$ and $q$ for larger values of $q$.

%Figure 3
%\begin{figure}
%\includegraphics [width=11.0cm] {Fig3aaJAP.eps}
%\includegraphics [width=11.0cm] {Fig3bbJAP.eps}
%\caption{(Color online) } \label{Fig3}
%\end{figure}
%

\subsection{Fowler-Nordheim plots}
\label{AFN}

\begin{table*}
   \centering
 \renewcommand{\arraystretch}{1.5}

 %\begin{tabular}{|c|}
 % \hline
 %$H=0.1$ \\ \hline

 % \end{tabular}
\caption{Results from Figs. \ref{Fig3}(a) and (b) for LAFEs with the local work function $\phi=3.5$eV, considering several values of $q$: the slopes of the ordinary $J_{M}$-$F_{M}$-type FN plots considering two regions [(1) and (2) - as identified in Fig. \ref{Fig3}] of FN plots; $S^{1}_{M}$ and $S^{2}_{M}$, obtained by performing a linear regression that considers the SN barrier function $\nu_{F}^{SN}$ when calculating $J_{M}$ and $J^{j}_{L}$ [see Eqs. (\ref{Eq4}) and (\ref{Eq5})]; $\omega_{1}$ and $\omega_{2}$ values extracted performing a linear regression using the data in Fig.\ref{Fig3}(b) considering two regions (1) and (2); values of $f^{extr}_{1}$ and $f^{extr}_{2}$ calculated using the Eq.(\ref{Eq3}) [See the text for more details] considering two regions (1) and (2).}

\begin{tabular}{|c|c|c|c|c|c|c|c|c|c|}

\hline
$q$ & $S^{1}_{M} (V/nm)$ & $S^{2}_{M} (V/nm)$ & $\omega_{1}$ & $\omega_{2}$ & $\gamma^{approx}_{C1}$ & $\gamma^{approx}_{C2}$ & $f^{extr}_{1}$ & $f^{extr}_{2}$  \\

 \hline
 \hline

$0.80$ & $-0.0788 \pm 0.00006$ & $-$ & $1.2179\pm 0.0006$ & $-$ & $656.64$ & $-$ & $0.26$ & $-$   \\ \hline
$0.61$ & $-0.0646 \pm 0.0002$ & $-0.10250\pm 0.00005$ & $1.012\pm0.003$ & $1.527 \pm 0.001$ & $665.57$ & $632.93$  & $0.18$ & $0.37$   \\ \hline
$0.56$  & $-0.06400 \pm 0.00005$ & $-0.11046 \pm 0.00005 $ & $1.003 \pm 0.001$ & $1.617 \pm 0.002$ & $665.83$ & $621.24$ & $0.19$ & $0.36$  \\ \hline\
$0.51$  & $-0.0642\pm 0.0001$ & $-0.12860 \pm 0.00004$  & $1.0015 \pm 0.0004$ & $1.761 \pm 0.002$ & $662.77$ & $581.78$ & $0.19$ & $0.38$  \\ \hline
$0.43$ & $-0.06351 \pm 0.00002$ & $-0.14300 \pm 0.00003$  & $1.00026 \pm 0.00007$ & $1.986 \pm 0.001$ & $669.14$ & $590.05$ & $0.24$ & $0.41$   \\ \hline

\end{tabular}
%\vspace{0.5cm}

\label{tab}
\end{table*}
%

%Figure 3
\begin{figure}
\includegraphics [width=10.0cm,height=7cm] {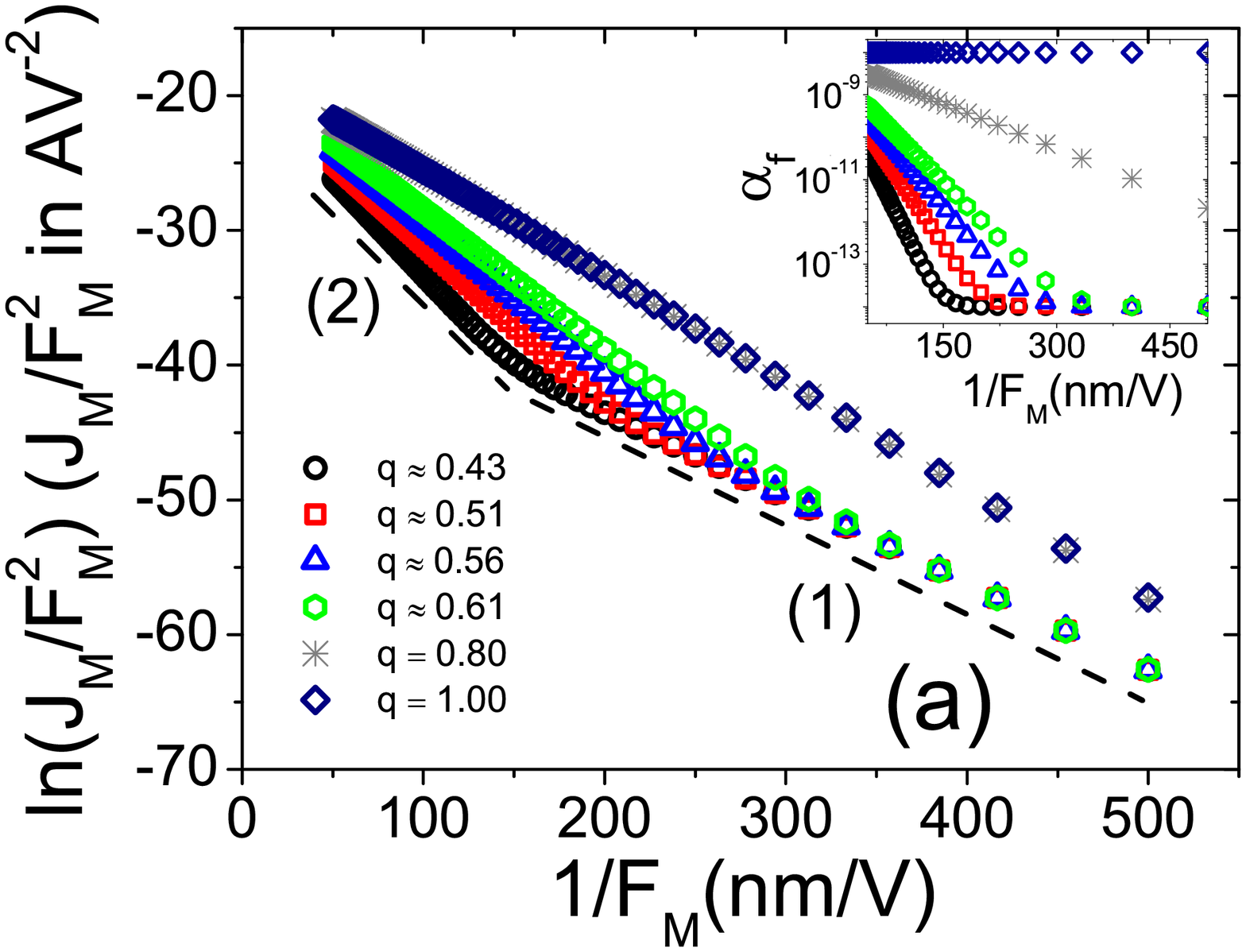}
\includegraphics [width=10.0cm,height=7cm] {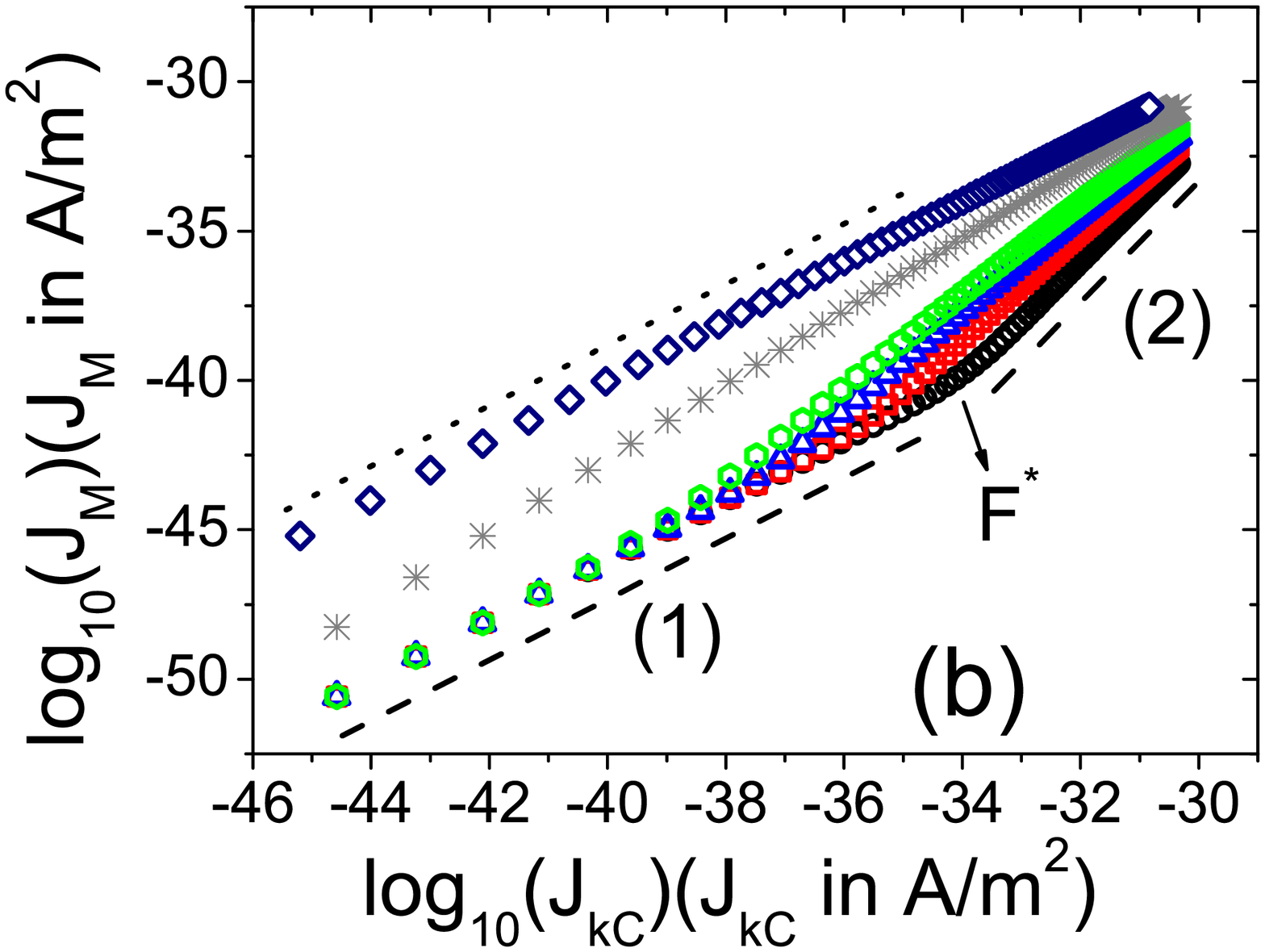}
\caption{(Color online) (a) Ordinary $J_{M}$-$F_{M}$-type FN plots for several values of $q$, $\rho(\gamma_{1}) = 10^{-6}$, and macroscopic electric field in the range of 2 V/ $\mu$m $\leq F_M \leq$ 20 V/$\mu$m [clear two slopes are highlighted in (1) ans (2)]. The data for an uniform LAFE with all local $\gamma = 552$ (q = 1.00) is also shown. In the inset, is shown the dependence between $\alpha_{f}$ and $1/F_{M}$. (b) Macroscopic current density, $J_{M}$ [see Eq. (\ref{Eq5})] as a function of the kernel characteristic current density, $J_{kC}$ [see Eq. (\ref{Eq7})] for the same parameters used in (a). The dashed lines show two quasi-linear sections (1) and (2) also verified in (a). The dotted line has slope 1 and is parallel to the dashed line  of section (1). The results are presented for $\Omega = 10^{-8}$ (see text for more details).} \label{Fig3}
\end{figure}
We now discuss the effect of the FEF distribution on the FN plots. Fig.\ref{Fig3}(a) presents FN plots for several values of $q$ and a fixed $\rho(\gamma_{1}) = 10^{-6}$, for the typical range of applied field  2 V/ $\mu$m $\leq F_M \leq$ 20 V/$\mu$m in CFE for vacuum nano-electronic technologies. It's possible to identify two well separated regions with a sharp crossover between two different slopes $S_{M}^{1}$ and $S_{M}^{2}$, when $q \in [0.43,0.61]$. In Table I, we list all pertinent values resulting from the analysis presented in Figs. \ref{Fig3}(a) and \ref{Fig3}(b). In the $q \rightarrow 1$ limit, the two slope pattern becomes less evident and linear behavior prevails. The inset of Fig.\ref{Fig3}(a) shows the behavior of $\alpha_{f}$ as a function of $1/F_{M}$, indicating that the nonlinear behavior on the FN plots is related to the dependency between $\alpha_{f}$ and $F_{M}$. In the low macroscopic electric field limit, it's possible to identify, for $q \lesssim 0.61$, that $\alpha_{f}$ presents a constant behavior, suggesting that the main emitting spots correspond to the regions with $\gamma_{j}=\gamma_{1}$. In the high $F_M$ limit, $\alpha_{f}$ depends exponentially on $1/F_{M}$, as expected from Eqs. (\ref{Eq11}) and (\ref{Eq12}). Here, the regions with $\gamma_{j}=\gamma_{2}$ contribute to the field electron emission.

Our results for the relation between the $J_M$ and $J_{kC}$ [see Eqs.  (\ref{Eq5}) and (\ref{Eq7})] add valuable insights to the discussion about the physical reasons that are responsible for the crossover phenomenon in FN plots. Previous works suggest that the weak nonlinear dependency in FN plots could be traced back to a simple relation $J_M$ to $J_{kC}$, namely $J_M \sim J_{kC}^{\omega}$, where $\omega$ has a weak dependency on $F_{M}$ but is strongly influenced by the LAFE geometry \cite{deAssis12015,deAssis22015}. This effect provides a more general method for a reliable assessment of the characteristic FEF $\gamma_{C}$ from FN plots. A good approximation $\gamma_{C}^{aprx}$ for the true FEF $\gamma_{C}$ was derived in \cite{deAssis12015,deAssis22015}, which leads to

\begin{equation}
\gamma_{C}^{aprx} = - \omega s_{t} b \phi^{3/2} / S_{M} = \omega s_{t} \beta^{FN},
\label{Eq13}
\end{equation}
where $s_{t}$ was introduced in Eq. (\ref{Eq3}). Under orthodox emission conditions the situation is that, if $\alpha_{f}$ does not depend on $F_{M}$, $\beta^{FN}$ generally over-predicts $\gamma_{C}$ by approximately 5\%. As anticipated in the Sec.\ref{sec.II}, $s_t \approx 0.95$ is verified for practical circumstances \cite{ForbesSlope}. The correction $\omega$, which was introduced very recently by one of authors \cite{deAssis12015, deAssis22015}, accounts for a nonlinear relationship between the macroscopic and the characteristic local current density, both of which are accessible experimentally.

In Fig.\ref{Fig3}(b), we illustrate the behavior of $J_M$ as a function of $J_{kC}$ for the same parameters used in Fig.\ref{Fig3}(a). We clearly identify that the same two slope patterns in the FN plots is observed for the dependency between $J_M$ and $J_{kC}$. Thus, it's convenient to define $\omega_{1}$ and $\omega_2$ so that

\begin{equation}
\gamma_{Cn}^{aprx} = - \omega_{n} s_{t} b \phi^{3/2} / S^{n}_{M} \hspace{1.0cm} (n=1,2),
\label{Eq14}
\end{equation}
where $\gamma_{C1}^{aprx}$ and $\gamma_{C2}^{aprx}$ correspond to the approximations for the characteristic FEF using the slopes $S^{1}_{M}$ and $S^{2}_{M}$, respectively. The results in Fig.\ref{Fig3}(b), together with Eqs. (\ref{Eq8})-(\ref{Eq12}), suggest that:

\begin{equation}
J_{M}  \sim J_{kC}^{\omega_{1}} \hspace{1.0cm} (F<F^{*}),
\label{Eq15}
\end{equation}
and

\begin{equation}
J_{M}  \sim J_{kC}^{\omega_{2}} \hspace{1.0cm} (F>F^{*}).
\label{Eq16}
\end{equation}
Here $F^{*}$ denotes the value of the electric field at the crossover point that separates the regions with two different slopes in FN plots as indicated in Fig.\ref{Fig3}(b). In Appendix - B, we provide detailed derivation of the expressions that allow to extract the parameter ``$r$" from similar nonlinear FN plots in orthodox CFE experiments. ``$r$" is a function of $F^{*}$, $S^{1}_{M}$, $S^{2}_{M}$ as well as of the local work function that through the exponent $\kappa$.

The results in Table I indicate that $\omega= \omega_{1} \approx 1.0$ in the low $F_M$ regime. The slope $S_{M}^{1}$ provides information on the characteristic FEF, $\gamma_{C}=\gamma_{1}$. In this regime, the results reinforce the interpretation that CFE is orthodox, as confirmed by the extracted value $f^{extr}_{1}$ [see Eq.(\ref{Eq3}) of this work, and Table 2 in Ref. \cite{Forbes1}, for $\phi=3.5$eV]. On the other hand, for high values of $F_{M}$, Table I indicates $\omega_{2}>1$, which means that, besides the regions with $\gamma_{j}=\gamma_{1}$, the regions with $\gamma_{j}=\gamma_{2}$ also contributes in a significant way to $\alpha_{f}$. This suggests an important result that might be suitable for experimental observation: when $\omega_{2}>1$ in the corresponding range of $F_{M}$, the slope $S^{2}_{M}$ provides information regarding the macroscopic FEF, $\gamma_{2} < \gamma_{C}$. A good estimate of the real characteristic FEF would be $\gamma_{C2}^{aprx} = - \omega_{2} s_{t} \beta^{FN}$, for $F_{M}>F^{*}$. For this ansatz, the errors do not exceed $15\%$, as indicated in Table I for $q \approx 0.43$. More interestingly, the values of $f^{extr}_{2}$ shown on Table I (extracted from the range $F_{M} > F^{*}$), confirm that the emission is also orthodox.

%changes
At this point, we emphasize the importance of measuring $\omega$. To see this, let us consider two different LAFE devices: (i) the first one is characterized  by uniform local FEFs with $\gamma_{1}=\gamma_{2}=552$ (and $q=1$); (ii) the second one is composed by regions with two distinct FEFs values, namely $\gamma_{1}=690$ and $\gamma_2=552$ ($q=0.8$) and $\rho({\gamma_{1}})=10^{-6}$. The device (i) represents an ideal homogeneous array composed by the same second-scale structures. Device (ii) represents an array where most of the second-scale structures are characterized by $\gamma_{j}=\gamma_{2}$, but there is a small probability to find regions with $\gamma_{j}=\gamma_{1}$, as already discussed in the characterization of a non-uniform LAFE surface. Both corresponding FN plots are shown in Fig. 3(a), but the two curves are actually indistinguishable. However, the results in the inset show that, while $\alpha_{f}$ is independent of $F_M$ in case (i), $\alpha_{f}$ does depend on $F_M$ for the device (ii). These observations culminate with the following conclusions: although FN plots present the same behavior for two distinct LAFE surfaces, in case (i) the corresponding slope provides the correct value of the characteristic FEF. On the other hand, the device (ii) has characteristic FEF $ \gamma_{C}=\gamma_{1}>\gamma_{2}$. Thus, the linear aspect of the FN plot does not mean, necessarily, that the area of emission does not depend on the macroscopic field. Indeed, the results in the inset of Fig. 3(a) for device  (ii) hints at change in the value of $\alpha_{f}$ by, at least, two orders of magnitude. Moreover, despite the linear aspect and the orthodox CFE, the FN slope can not measure, necessarily, the characteristic FEF, $\gamma_{C}$.  This reflects the importance of measure $\omega_{n}$, so that $\omega_{n} > 1$ suggests this behavior. Finally, we remark that if $\omega_{n} \approx 1$ for a given $F_{M}$ range in CFE  experiments, it just indicates that $\alpha_{f}$ does not depends (or weakly depends) on the $F_{M}$ in that range.

\subsection{Application to Isolated Nanopost Field Emitters (with $\Omega \approx 10^{-7}$)}
\label{Posts}

\begin{figure}
\centering
\includegraphics [width=12.0cm] {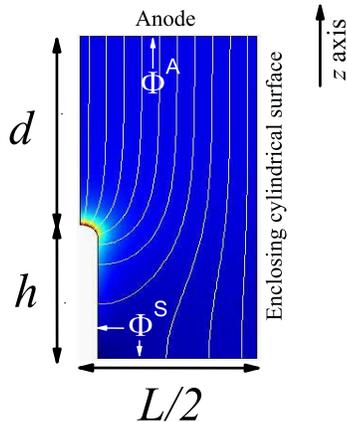}
\caption{(Color online) Two dimensional representation of a tip placed in the center of a $L \times L$ location used in the simulations. Parameters $h$, $d$ and $L$  represent the height of a nano-emitter, the distance from its apex to the far away anode, and the half of the lateral size of the integration domain, respectively. $\Phi^{S}$ and $\Phi^{A}$ indicate, respectively, the fixed electric potential of the emitter and of the anode, as required by the Dirichlet conditions used in numerical simulations.
The electric field lines and the enclosing cylindrical surface are also shown. The macroscopic electric field component, perpendicular to the displayed region, vanishes identically.
The emitter may experiences a screening effect due to its own image, similar to the screening in a lattice. In this work we use $L=5h_{1}$ (see text for more details), so that the screening is negligible. For the purpose of calculating area efficiencies, we assume that each post-like emitter has a ``footprint" of area $L^{2}$.} \label{Fig40}
\end{figure}

\label{laplace}
\begin{figure}
\centering
\includegraphics [width=8.0cm] {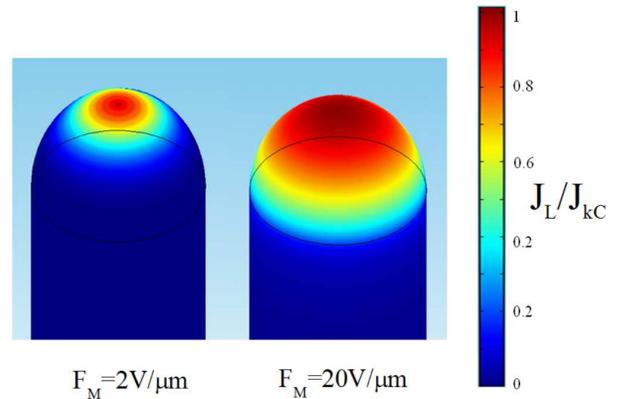}
\caption{(Color online) Normalized local current density map ($J_{L}/J_{kC}$) for emitter with $\gamma_{C}=\gamma_{1}=678$ at macroscopic electric fields $2V/\mu$m and $20V/\mu$m.} \label{Fig40a}
\end{figure}

%Figure 4
\begin{figure}
\includegraphics [width=10.0cm,height=7cm] {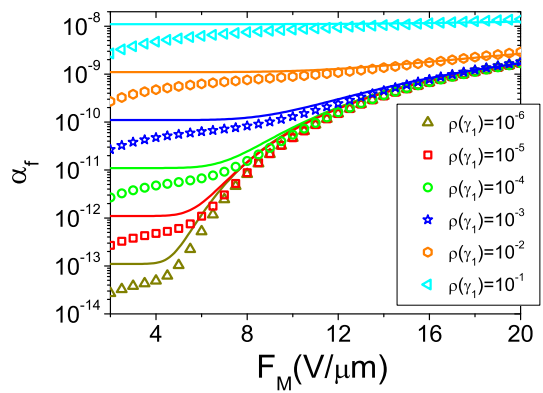}
\caption{(Color online) Comparison between $\alpha_{f}$ as a function of $F_{M}$ for two different conditions. Solid lines indicate the solutions obtained from Eqs.(\ref{Eq8}-\ref{Eq12}), for $\Omega\approx10^{-7}$, while hollow symbols indicate the results from numerical solution of Laplace's equation.} \label{Fig5}
\end{figure}
%

%Figure 5
\begin{figure}
\includegraphics [width=10.0cm,height=7cm] {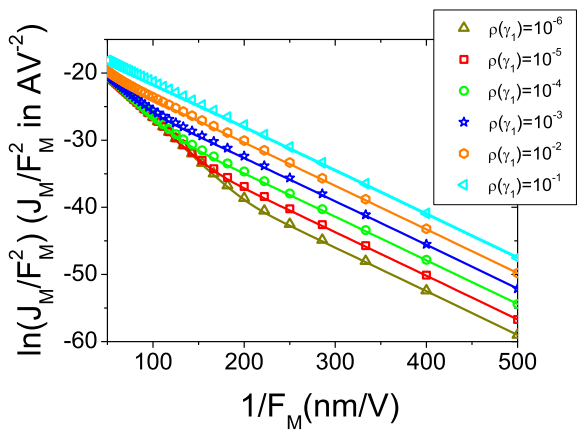}
\caption{(Color online) Comparison of the FN plots for the same conditions shown in Fig. \ref{Fig5}. Solid lines indicate results obtained from Eqs.(\ref{Eq8}-\ref{Eq12}), for $\Omega\approx10^{-7}$, while hollow symbols correspond to the numerical solution of Laplace's equation.} \label{Fig6}
\end{figure}

In this section, the validity of the former analysis is compared with those for a structured emitter. We assume the single emitters as structures shown in Fig.\ref{Fig40}, which are usual representations of nano-emitters as a hemisphere on a conducting cylindrical post \cite{CNT1,CNT2,Cole,MC}. We solve numerically the Laplace's equation, in a three dimensional domain, using an array of conducting nano-emitters with large apex radii ($R=50$ nm) but different heights, $h_{1}$ and $h_{2}$ ($h_{1} > h_{2}$), which are associated to the FEFs $\gamma_{1}$ and $\gamma_{2}$, respectively. In our analysis, we fix $q\approx 0.51$, with $\gamma_{1}=678$ and $\gamma_{2}=346$. This corresponds, in our simulations, to nanostructures with aspect ratios ($h/R$) close to $1193$ and $555$, respectively. The latter are compatible with field emission displays where electrons are emitted from micron-sized tips \cite{PatFE}. The electric potential distribution on the integration domain was calculated using a Finite Element Method scheme (software COMSOL® v4.3b). This allows to calculate the electric field distribution over the LAFE device, as well as the local emitting current density using Eq.(\ref{Eq4}). We consider the same work function, $\phi=3.5$eV used in the previous section. Fig. \ref{Fig40} shows the radial integration domain (emitting location) and the used boundary conditions for an idealized situation in which a single tip is placed in the center of a $L \times L$ location. The line at the right side boundary generates an enclosing cylindrical surface (ECS) when it is rotated by $2 \pi$ around the position where the left boundary lies. In this way, the electric field component normal to this plane is locally zero everywhere. Since a similar geometry may be found in the neighboring locations, with the exception that the tips do not necessarily lie in the corresponding location centers, the resulting field may be distorted as a consequence of the superposition of individual field at each location.  Thus, there is an overall screening effect inside each ECS. In this work we use $L=5 h_{1}$ and $d=\sqrt{2}L$, so that the screening is negligible (the emitters can be considered as isolated) and the field lines can be considered parallel and vertically aligned \cite{FAD}. The electric potential $\Phi^{A}\neq 0$ of the anode at the top boundary guarantees electric field intensity equal to $F_M$ at the boundary. Moreover, the emitter surface and the bottom boundary of the cell are grounded ($\Phi^{S}$=0). For the purpose of calculating area efficiencies, we assume that each post-like emitter has ``footprints" of area $L^{2}$.

The macroscopic current density was calculated as follow:

\begin{equation}
J_{M} = \frac{1}{L^{2}}\left\{\rho{(\gamma_{1})} \sum\limits_{cap} J_{L}^{1} \Omega\Delta A_{L}^{1} + \rho{(\gamma_{2})} \sum\limits_{cap} J_{L}^{2} \Omega\Delta A_{L}^{2}\right\},
\end{equation}
where the sum is computed over all spherical cap surface area and $\rho{(\gamma_{1})}$ and $\rho{(\gamma_{2})}$ correspond to probabilities to found a location of LAFE that contains a nanostructure with characteristic FEF $\gamma_{1}$ and $\gamma_{2}$, respectively.
In this case, $\alpha_{f}$ may changes essentially for two reasons: (i) the emitters with FEFs $\gamma_{2}$  contribute to the overall current; (ii) the notional area on each emitter increases slowly as $F_{M}$ increases, as shown in Fig.\ref{Fig40a}. To illustrate this dependency, we have computed the normalized local current density map ($J_{L}/J_{kC}$) at macroscopic electric fields 2V/$\mu$m and 20V/$\mu$m. In fact, it is possible to observe a clear increase of the notional area of a single nano-emitter, as first suggested by Abott and Henderson \cite{Abott} in 1939. In Fig.\ref{Fig5}, we show a comparison for the dependency of $\alpha_{f}$ as a function of $F_{M}$ for two methodologies: the one based on Eqs.(\ref{Eq11}) and (\ref{Eq12}), and that obtained by solving Laplace's equation. In the latter, using the dimensions previously discussed, $\Omega \sim R^{2}/L^{2} \sim 10^{-7}$. Moreover, the results suggest that $\Omega \Delta A_{L}^{j}$ is weakly dependent on $F_{M}$. Then, in Eq.(\ref{Eq5}) we have used the reasonable proportionality $\Omega \Delta A_{L}^{j} \sim \pi R^{2}$, which means to use $\Omega \approx 10^{-7}$ in Eq.(\ref{Eq11}). It's possible to observe the good agreement between two results. A small deviation occurs in low $F_{M}$ regime, which can be justified because the emitting area of a single tip structure grows very slowly as the macroscopic electric field increases (see Fig.\ref{Fig40a}). However, an important result is that this very subtle effect does not affect the form of FN plots. Fig.\ref{Fig6} shows the nonlinear behavior of FN plots for actual emitters, considering $10^{-6} \leq \rho(\gamma_{1}) \leq 10^{-1}$ and $q=0.51$, showing the excellent agreement with the results from Eqs.(\ref{Eq11}) and (\ref{Eq12}).

\section{Conclusions}
\label{conc}

In this work, we present a theoretical explanation for the crossover in the behavior of the FN plots, commonly found for large area field emitters with irregular morphology. The latter is assumed to lead to a more prominent emitting locations with FEFs distributed approximately as a bimodal distribution. Our results suggest an orthodox field electron emission for two quasi-linear sections of FN plots as the formal area efficiency is the sole cause of the crossover, in a typical range $F_M \in $ $[2,20]$ V/$\mu$m. For such situations, we propose a physically relevant ansatz leading to the interpretation of the slopes in FN plots as a function of the $q$ and $r$ asymmetry parameters characterizing $\rho(\gamma)$. Finally, the results from solution of Laplace's equation for an array of conducting nano-emitters supports our theoretical assumptions regarding the information provided by FN plots, which can be tested if CFE experiments are orthodox.

\section*{Acknowledgements}

The authors acknowledge the financial support of the Brazilian agency CNPq. TAdA thanks R. G. Forbes for fruitful discussions and for calling attention to Ref.[6].

\vspace{1.0cm}

\section*{APPENDIX}

\subsection{Derivation of $\alpha_{f}$}

According to Eqs.(\ref{Eq5}) and (\ref{Eq9rev}), the macroscopic current density for a LAFE with two prominent emitter locations can be written as

\begin{equation}
J_M = \frac{n_L}{A_M} \left\{\rho(\gamma_1)J^{1}_L \Omega\Delta A^{1}_L + \rho(\gamma_2) r J^{2}_L \Omega\Delta A^{2}_L\right\}.
\label{AP1}
\end{equation}
We emphasize that, in our theory, $\Delta A^{j}_L$ represents the footprint area of $j-$th post-like emitter. $\Omega\Delta A_L$ represents the corresponding notional emission area. Then, using Eq.(\ref{Eq4}) (for $\lambda_{L}=1$), assuming that $\Omega$ is weakly field dependent, and $\Delta A^{1}_L = \Delta A^{2}_L = \Delta A_L$, Eq.(\ref{AP1}) becomes

\begin{multline}
J_M = \frac{n_L \Omega\Delta A_L}{A_M} \rho(\gamma_1) [ (\gamma_{1}F_{M})^{(2-\eta/6)} \exp\{-b\phi^{3/2}/\gamma_{1}F_M\} + \\
 + r(q\gamma_1 F_{M})^{(2-\eta/6)} \exp\{-b\phi^{3/2}/q\gamma_{1}F_M\} ].
\label{AP2}
\end{multline}

Once the term $\exp\{-b\phi^{3/2}/\gamma_{1}F_M\}$ appears in both terms, we take into account that $n_{L}\Delta A_L = A_{M}$, to simplify Eq.(\ref{AP2}) to

\begin{multline}
J_M = \Omega\rho(\gamma_{1}) \times \\
\times \left\{1 + q^{\kappa} r \exp\left[-b(q^{-1} - 1)\phi^{3/2}/(\gamma_{1}F_M)\right]\right\} J_{kC},
\label{AP3}
\end{multline}
where $J_{kC}$ is given by Eq.(\ref{Eq7}). Then, making use of the notation introduced in Eq.(\ref{Eq8}), the formal area efficiency can be given by:

\begin{multline}
\alpha_{f}(F_M) \equiv \Omega\rho(\gamma_{1}) \left\{1 + q^{\kappa} r \exp\left[-b(q^{-1} - 1)\phi^{3/2}/(\gamma_{1}F_M)\right]\right\} \equiv \\
\equiv \Omega \rho(\gamma_1) \left\{ 1 + \Gamma{(q,r,\phi, F_{M})}\right\}.
\label{AP4}
\end{multline}

A generalization of Eq.(\ref{AP4}) that consider a LAFE with a larger number of tips types, i.e. with $\{\gamma_{j}\}$ (j=1,...,n), can be easily derived, leading to

\begin{multline}
\alpha_{f}(F_M)  \equiv \Omega\rho(\gamma_{1}) \sum\limits_{j=1}^{n} q_{j}^{\kappa} r_{j} \exp\left[-b(q_{j}^{-1} - 1)\phi^{3/2}/(\gamma_{1}F_M)\right],
\label{AP5}
\end{multline}
where $q_j = \gamma_j / \gamma_1$ and $r_j = \rho(\gamma_j)/\rho(\gamma_1)$.

\vspace{0.5cm}

\subsection{Extraction of parameter ``$r$" from nonlinear FN plots in orthodox CFE experiments}

If CFE experiments are orthodox and the FN plots present two clear-cut quasi-linear sections, it's possible to provide an estimation of the parameter ``$r$" defined in Eq.(\ref{Eq9rev}). Let the macroscopic electric field at the crossover point that separates the regions with two different slopes be noted by $F^{*}$, as illustrated in Fig.\ref{Fig3}(b). At this point, it is expected that the contribution for macroscopic current density from the locations with FEF $\gamma_1$ is the same as those from the locations with FEF $\gamma_2$. This lead to

\begin{multline}
\rho(\gamma_1) (\gamma_1 F^{*})^{\kappa} \exp{[-b\phi^{3/2}/(\gamma_1 F^{*})]} = \\
= \rho(\gamma_2) (\gamma_2 F^{*})^{\kappa} \exp{[-b\phi^{3/2}/(\gamma_2 F^{*})]}.
\label{AP21}
\end{multline}
From Eq.(\ref{AP21}), it's possible to write the product $rq^{\kappa}$ as

\begin{equation}
rq^{\kappa} = g(F^{*})^{\left(\frac{1}{\gamma_1} - \frac{1}{\gamma_2}\right)},
\label{AP22}
\end{equation}
where $g(F^{*}) \equiv \exp{[-b\phi^{3/2}/F^{*}]}$. From the expressions for the two distinct slopes in the same corresponding FN plot, $\gamma_1 = - s_{t} b \phi^{3/2}/S^{1}_M$ and $\gamma_2 = - s_{t} b \phi^{3/2}/S^{2}_M$, it's possible to write

\begin{equation}
\frac{1}{\gamma_1} - \frac{1}{\gamma_2} = -\frac{1}{s_{t}b\phi^{3/2}} \left(S^{1}_M - S^{2}_M\right).
\label{AP23}
\end{equation}

Finally, using Eqs.(\ref{AP22}) and (\ref{AP23}), the parameter $r$ is given by:

\begin{equation}
r=\exp{\left[\frac{(S^{1}_M - S^{2}_M)}{s_{t}F^{*}}\right]} \left[\frac{S^{1}_M}{S^{2}_M}\right]^{-\kappa}.
\label{AP24}
\end{equation}
%

%~~~~~~~~~~~~~~~~~~~~~~~~~~~~~~~~~~~~~~~~~~~~~~~~~~~~~~~~~~~~~~~~~~~~~~~~~~~
%~~~~~~~~~~~~~~~~~~~ REFERENCES ~~~~~~~~~~~~~~~~~~~~~~~~~~~~~~~~~~~~~~~~~~
%~~~~~~~~~~~~~~~~~~~~~~~~~~~~~~~~~~~~~~~~~~~~~~~~~~~~~~~~~~~~~~~~~~~~~~~~~~~
%

\end{document}